%% file: main.tex
\documentclass[journal]{IEEEtran}
\usepackage{graphicx}
\usepackage{svg}
\usepackage{color}
\usepackage{cite}
\usepackage{amsmath,amssymb}
\usepackage{threeparttable}
\usepackage{matlab-prettifier}
\usepackage{pgfplots}
\pgfplotsset{compat=1.9}
\usepackage{tabularx,booktabs}
\usepackage{wasysym}
\usepackage[bookmarks=false]{hyperref} 
\usepackage{float}
\newcolumntype{C}{>{\centering\arraybackslash}X} 
\usepackage{siunitx}
\usepackage{caption}
\usepackage{circuitikz}
\usepackage{tikz}
\usepackage{tikzscale}
\usepackage{derivative}

\usepackage{dblfloatfix}
\usepackage{array}
\usepackage{tabularx}
\usepackage{booktabs}      
\usepackage{xcolor}        
\usepackage{makecell}

\usepackage{color}

\begin{document}

\title{Broadband 600- and 1200-GHz Subharmonically Pumped Schottky Diode Mixers}
\author{Peter J.~Sobis,~
    Vladimir~Drakinskiy,~\IEEEmembership{Member,~IEEE,}
    Karl~Jacob,~
    Axel~Murk,~
    Anders~Emrich,~
    and~Jan~Stake,~\IEEEmembership{Fellow,~IEEE}

\thanks{\noindent Manuscript received October 11\textsuperscript{th}, 2026; revised xxxx. This research was in part funded by the Swedish National Space Agency (SNSA), European Space Agency (ESA) under contract 4000112396/14/NL/BW "Terahertz Receiver Technology for Future Science Missions", Swiss National Science Foundation under Grant 200020-165744, and was carried out in the Advanced Digitalization program at the WiTECH Center CRYTER+, which has been financed by VINNOVA, Chalmers University of Technology, Omnisys Instruments AB (AAC Omnisys), Low Noise Factory AB (LNF), Research Institutes of Sweden (RISE) and Virginia Diodes Inc. (VDI).}
\thanks{Peter Sobis was with Omnisys Instruments AB and Chalmers University of Technology, SE-412 96 Gothenburg, Sweden. He is now with Low Noise Factory AB, Gothenburg, Sweden.}
\thanks{Karl Jacob was with the Institute of Applied Physics, University of Bern, Bern, Switzerland.}
\thanks{Axel Murk is with the Institute of Applied Physics, University of Bern, Bern, Switzerland.}
\thanks{Anders Emrich was with Omnisys Instruments AB. He is now with EmriX Consulting AB, Askim, Sweden.}
\thanks{Vladimir Drakinskiy and Jan Stake are with the Department of Microtechnology and Nanoscience, Chalmers University of Technology, SE-412 96 Gothenburg, Sweden. (e-mail: \mbox{jan.stake@chalmers.se})}
}

\markboth{Submitted to IEEE TRANSACTIONS ON TERAHERTZ SCIENCE AND TECHNOLOGY / THz Letters, 2026}%
{Stake \MakeLowercase{\textit{et al.}}}


\maketitle

\begin{abstract}
A subharmonic mixer topology is introduced that uses quarter-wave stubs to create virtual RF/LO grounds and provide impedance matching in close proximity to the antiparallel Schottky barrier diode pair, thereby reducing the electrical length and enabling broadband operation. The RF/LO matching topology is demonstrated at 600 GHz and 1200~GHz, with good agreement between measured and simulated performance. The 600-GHz mixer demonstrates approximately 20\% RF bandwidth with a DSB receiver noise temperature below 1300~K across most of the measured band, while the 1200-GHz implementation achieves DSB receiver noise temperatures below 3000~K over a broad frequency range. Both mixers are characterized at room temperature with approximately 2~mW of LO power. The 600-GHz mixer is used in the Submillimeter Wave Instrument (SWI) onboard ESA’s Jupiter Icy Moons Explorer. Overall, these results demonstrate state-of-the-art room-temperature performance for broadband receiver front ends targeting atmospheric research and remote-sensing applications.
\end{abstract}


\input{Chapter/intro}

\input{Chapter/method}

\input{Chapter/results}

\section*{Acknowledgment}
The authors would like to thank the SWI team, led by principal investigator Paul Hartogh at the Max Planck Institute for Solar System Research (MPS, Germany), and Niklas Wadefalk at Low Noise Factory for the LNA design. We would also like to thank Olivier Auriacombe and Mats Lindgren at AAC Omnisys in Sweden, and Divya Jayasankar at DLR in Germany, for their valuable input, encouragement, and feedback throughout the drafting process.

The integrated Schottky barrier diode circuits were fabricated and measured in the Nanofabrication Laboratory (MyFab Chalmers) and the Kollberg Laboratory at Chalmers University of Technology in Gothenburg, Sweden.

\bibliographystyle{IEEEtran}
\bibliography{IEEEfull,bibl}

\end{document}

%% file: Chapter/intro.tex
\section{Introduction}
\IEEEPARstart{S}{ubmillimeter}-wave heterodyne receivers \cite{Hubers2008} are crucial for high-resolution remote sensing, especially in space applications \cite{Siegel2007}, where they are used to gather meteorological data and observe planetary and comet atmospheres \cite{Thomas2025}. Successful space missions like Odin \cite{Murtagh2002}, NASA's EOS MLS \cite{Waters2006}, MIRO \cite{Gulkis2006}, and the Herschel space observatory \cite{deGraauw2010} span millimeter and submillimeter wavelengths, providing scientific observations. Superconducting receivers offer high sensitivity but require cooling \cite{Zmuidzinas2004}, limiting their long-term use in space. In addition, a wide instantaneous bandwidth is essential for quick, broad spectral data collection. Therefore, developing low-noise, compact, room-temperature or passively cooled receivers is necessary for future space science missions. For this, Schottky-based technology is vital, as discussed in the review \cite{Mehdi2017}. 

Since the initial development of fixed-tuned submillimeter-wavelength mixers employing planar Schottky barrier diode (SBD) technology \cite{Bishop1987}, as demonstrated by Hesler \textit{et al.} in 1997 \cite{Hesler1997}, substantial advancements have been achieved in integrated diode technology, including the membrane-integrated 2.5-THz mixer \cite{Siegel1999} developed at JPL and the 520-590-GHz biased subharmonic mixer topology proposed by Schlecht \textit{et al.} in 2007 \cite{Schlecht2007}. Moreover, a subharmonic \mbox{1200-GHz} receiver with a double-sideband (DSB) mixer noise temperature of 2800~K was demonstrated in 2014 \cite{Schlecht2014}, and a noise temperature of 4000-6000~K was recently achieved at 2~THz \cite{Maestrini2025}.  Similarly, outstanding mixer performance was demonstrated with the LERMA-LPN Schottky process, achieving a noise temperature of 1070-1500~K between 520-620~GHz \cite{Treuttel2016} and about 4000~K between 1080-1280~GHz \cite{Treuttel2023} at room temperature. For the latter, operating at 120-150~K ambient temperatures, Treuttel \textit{et al.} measured a receiver noise temperature between 2400~K and 3000~K \cite{Treuttel2023}. Moreover, state-of-the-art mixer performance has been demonstrated using the Chalmers integrated Schottky diode process, which includes an optimum DSB receiver noise temperature of 450~K at 183~GHz \cite{Anderberg2019}, and 1300~K at 557~GHz \cite{HuanZhao2012}. However, the standard planar mixer topology \cite{Hesler1997} used in previous research poses several challenges for wide RF bandwidth operation, primarily due to electrically large matching circuitry. 

In this work, we present a broadband subharmonic mixer design that is demonstrated at 600~GHz and 1200~GHz. The circuit topology employs four open and short-circuited stubs to provide matching and virtual RF/LO grounds, respectively, in close proximity to the antiparallel Schottky diode pair, thereby reducing electrical length and facilitating wideband operation. This is possible thanks to advanced nanolithography and fabrication techniques, which enable precise control over the formation of airbridges, anodes, and ohmic contact pads.  The mixer was originally developed to meet the requirements of the Submillimeter Wave Instrument (SWI), a dual-channel heterodyne spectrometer, for the Juice (Jupiter Icy Moons Explorer) mission, an ESA mission launched on April 14, 2023. The receiver performance has been previously presented at conferences and workshops (600~GHz \cite{Sobis2015, Jacob2015, Jacob2019} and 1200~GHz \cite{Sobis2016}); however, this paper introduces the novel mixer topology and elaborates on its design details.

%% file: Chapter/method.tex
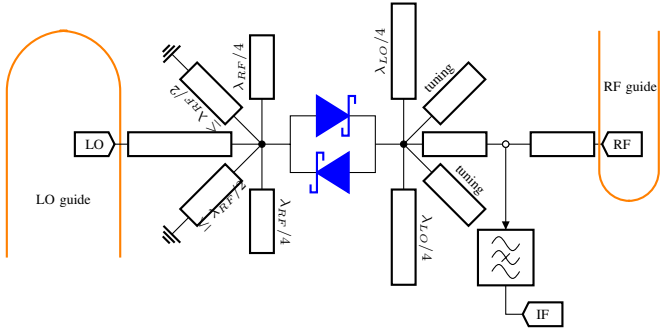
\begin{figure}[t]
    \centering
    \input{figs/fig1}
    \caption{Circuit topology. Equivalent circuit of the subharmonically pumped mixer. Four balanced stubs provide impedance matching and virtual ground for RF and LO frequencies.}
    \label{fig:schematic}
\end{figure}

\section{Method}

\subsection{Design}
Fig. \ref{fig:schematic} illustrates the mixer topology, with the antiparallel diode pair (APDP) connected in series between the RF and LO lines. Conventional \textit{E}-field probes are used to transition between the hollow waveguide and the suspended stripline modes. An open-ended $\lambda_{LO}/4$-balanced stub provides the virtual LO-ground, while an open-circuited $\lambda_{RF}/4$ balanced stub provides the virtual RF-ground. Both are located on the reference plane of the diode pair, resulting in an electrically compact circuit. A similar mixer topology has been demonstrated using an integrated coplanar-waveguide-based subharmonic mixer \cite{Chattopadhyay2011, Chattopadhyay2014patent}. In this work, additional balanced tuning stubs, short-circuited $\leq \lambda_{RF}/2$ and open-ended $\leq \lambda_{RF}/4$, are also implemented to facilitate IF ground and LO and RF impedance matching. The mixers were designed using finite-element 3D electromagnetic simulation of the circuit, combined with harmonic-balance analysis to analyze the overall mixer response. Both mixers utilize SBDs with an $n$-doping concentration of $3\times 10^{17}$~cm$^{-3}$, and anode diameters of 0.87/0.40~$\mu$m. A standard nonlinear diode model with an estimated junction capacitance of 1.3/0.3~fF \cite{Louhi1994} was used to analyze the 600-GHz and 1200-GHz mixers, respectively.   

The 1200-GHz mixer features an integrated conical RF-horn antenna, while the 600-GHz module uses an external horn. The 1200-GHz conical horn antennas were designed to meet the SWI optical requirements \cite{Kotiranta2018} and are based on a smooth-walled spline profile as described by Granet \textit{et al.} \cite{Granet_2004}. 

An integrated 4-13-GHz low-noise amplifier (LNF-LN4-13C) for the downconverted intermediate frequency (IF) is housed in the same waveguide package, thereby improving overall receiver noise performance \cite{Sobis2012}.

\begin{figure}[ht]
    \centering
    \includegraphics[width=0.9\linewidth]{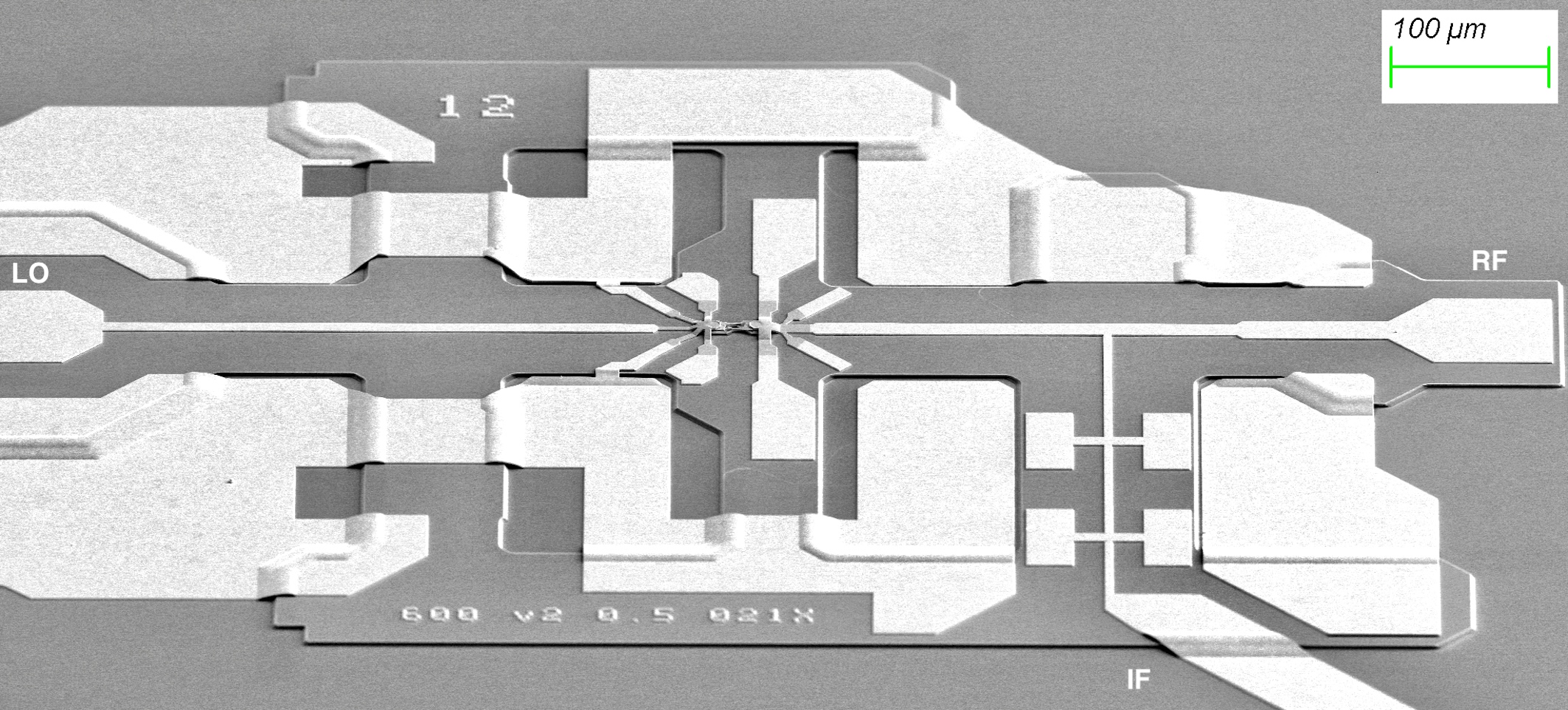}
    \caption{Integrated 600-GHz mixer circuit. SEM image of the integrated 600-GHz subharmonic mixer, showing the antiparallel GaAs Schottky diode pair with RF/LO matching circuit elements. The diameter of the anodes is 0.87 $\mu$m.}
    \label{fig:sem600}
\end{figure}

\begin{figure}[ht]
    \centering
    \includegraphics[width=0.9\linewidth]{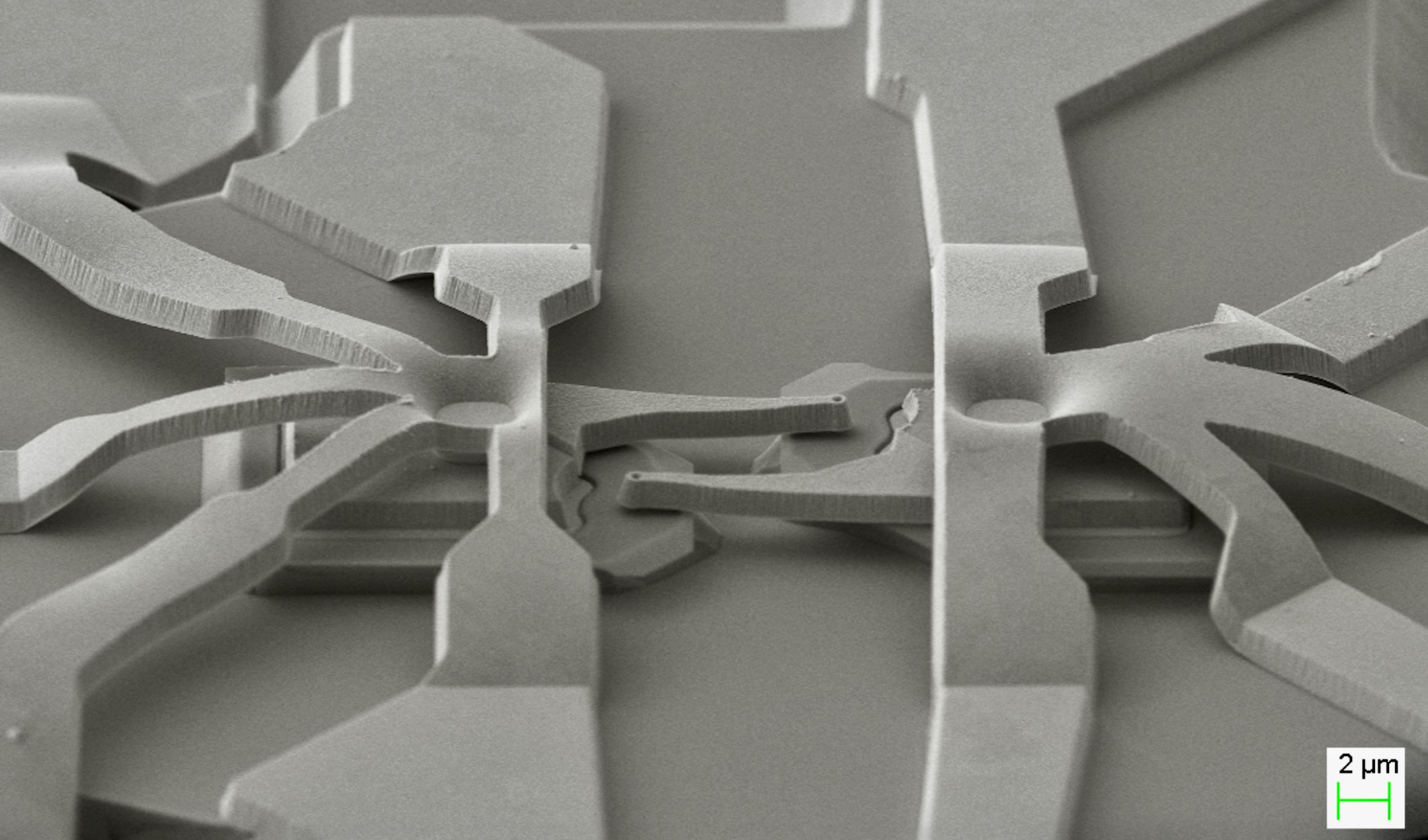}
    \caption{Integrated 1200-GHz mixer circuit. Close-up SEM image of the antiparallel GaAs Schottky diode pair. The diameter of the anodes is 400 nm.}
    \label{fig:sem1200}
\end{figure}

\subsection{Fabrication and assembly}
The integrated GaAs Schottky barrier diode circuits were fabricated at Chalmers University of Technology using electron-beam lithography, thereby enabling precise and repeatable formation of small anodes and air-bridges. The epitaxial structure comprises a 64-nm-thick cathode layer with an $n$-doping concentration of $3\times 10^{17}$~cm$^{-3}$, grown atop a 1.5-$\mu$m-thick, highly doped ($n^{++}$) contact layer. The device stack is supported by a 3-$\mu$m-thick undoped GaAs membrane. Fabrication commences with the protection of the device layer by a stress-balanced $\mathrm{SiO}_2$ layer deposited via PECVD. Ohmic contacts are established through wet etching, metal deposition, lift-off, and annealing. Subsequently, submicron anode contacts are fabricated by wet etching through the $\mathrm{SiO_2}$ layer, followed by the deposition of Schottky metallization and lift-off procedures. After visual inspection and size verification of the anodes via scanning electron microscopy (SEM), air-bridge connections and diode isolation are completed. The membrane is formed through selective wet etching of the 3-$\mu$m GaAs layer, which ceases at the underlying (Al,Ga)As layer. Passive circuitry, including beam leads, waveguide probes, and filter structures, is produced by metallization deposition and subsequent lift-off. Finally, the sample is thinned from the backside up to the (Al,Ga)As layer, which is then etched away to release the integrated diode circuits. Electrical current-voltage characterization of the diodes is performed through on-wafer measurements. The dc-series resistance and ideality factor for the mixers at 600~GHz and 1200~GHz are measured to be $R_s = 18/28\Omega$  and $\eta=1.18$, respectively. SEM images of the integrated circuit and antiparallel Schottky barrier diode pair are depicted in Figs. \ref{fig:sem600} and \ref{fig:sem1200}, respectively. The \textit{E}-plane waveguide split housings were machined from a single block of aluminum (Al 6082) alloy using a CNC machine and coated with a layer of electroplated gold over a nickel underplate.


\begin{figure}[h]
    \centering
    \includegraphics[width=0.9\linewidth]{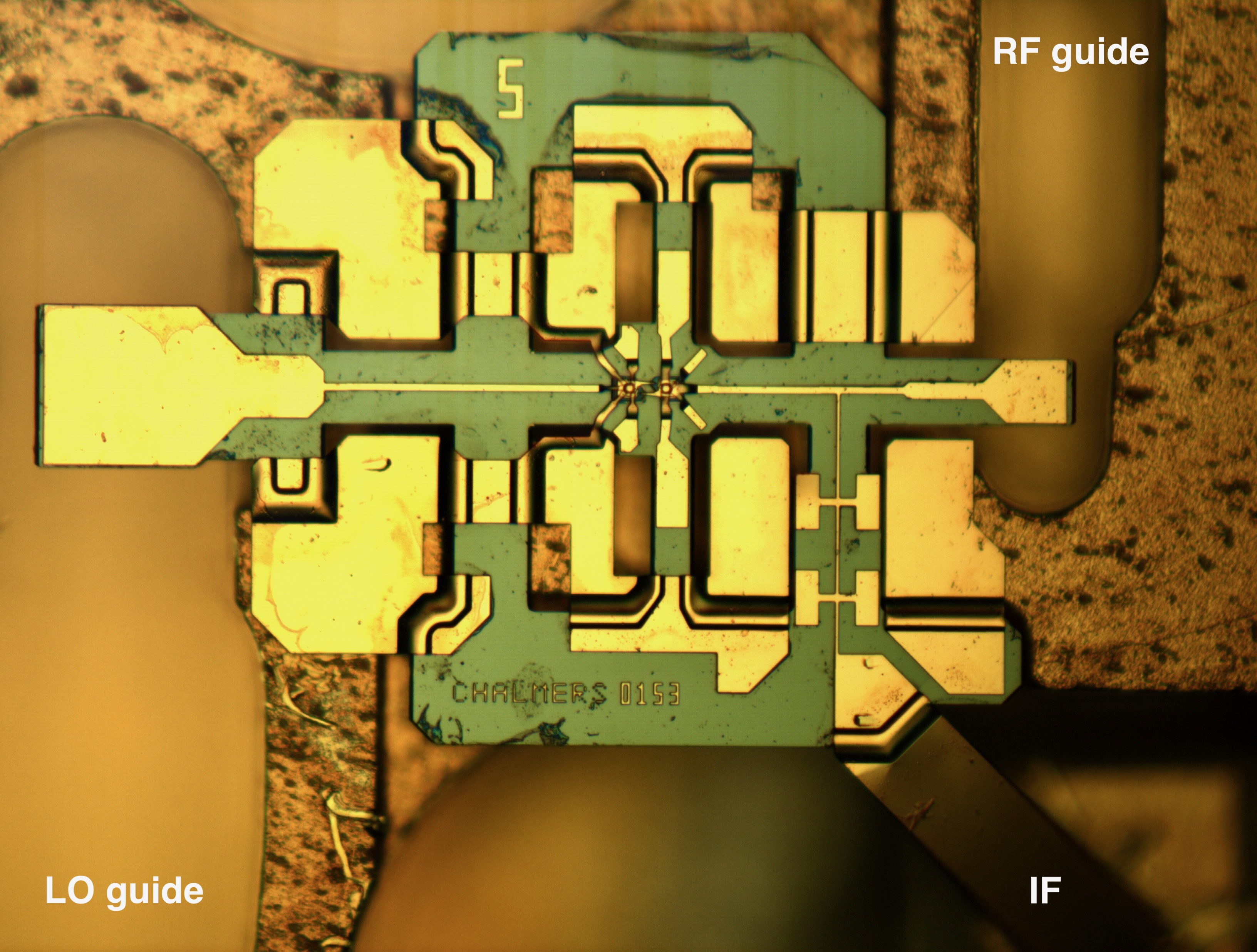}
    \caption{Assembly. Integrated 1200-GHz mixer circuit mounted in \textit{E}-plane split waveguide housing.}
    \label{fig:assembly}
\end{figure}

\subsection{Measurement setup}
The 1200-GHz receiver noise temperature was measured at Omnisys Instruments using the Y-factor method with at least one ambient and one thermally regulated hot conical load from TK Instruments. The LO consists of an E-band power amplifier (PA) from RPG, followed by three Schottky varactor doublers (Chalmers technology). The 600-GHz mixer was characterized at the University of Bern \cite{Jacob2015} using a conical ambient load and a LN\textsubscript{2} cold load with a TK-RAM absorber. The LO chain consisted of an E-band PA, a 140-GHz doubler from RPG, and a 280-GHz doubler from LERMA, and the  RF signal was received with a corrugated horn.

%% file: figs/fig1.tex
\begin{circuitikz}[
    scale=0.75,
    transform shape,
    every node/.style={font=\scriptsize}
]

\draw[rounded corners=0.8cm, thick, orange] (-5,-2) -- (-5,2) -- (-3,2) -- (-3,-2); 
\node at (-4,-1) {LO guide};

\draw (-3.1,0) node[msport, xscale=-1]{\ctikzflipx{LO}}
      to[mstline, mstlinelen=1.3] (-0.8,0)
      to[short] (-0.5,0) node[circ] (RF_short) {};

\draw (RF_short) to (-1,0.5) node[mslstub, mstlinelen=0.8, rotate=135](s1){$\leq \lambda_{RF}/2$};
\draw (s1.right) node[ground, rotate=225]{};

\draw (RF_short) to (-0.5,0.8) node[mslstub, mstlinelen=0.8, rotate=90](){$\lambda_{RF}/4$};

\draw (RF_short) to (-1,-0.5) node[mslstub, mstlinelen=0.8, rotate=-135](s2){\ctikzflipxy{$\leq \lambda_{RF}/2$}};
\draw (s2.right) node[ground, rotate=315]{};

\draw (RF_short) to (-0.5,-0.8) node[mslstub, mstlinelen=0.8, rotate=-90](){$\lambda_{RF}/4$};

\draw (RF_short) to[short] (0,0) -- (0,0.5)
      to[full Schottky diode, color=blue] (1.5,0.5) -- (1.5,-0.5)
      to[full Schottky diode, color=blue] (0,-0.5) -- (0,0);

\draw (1.5,0) to[short] (2,0) node[circ] (LO_short) {};

\draw (LO_short) to (2.5,0.5) node[mslstub, mstlinelen=0.8, rotate=45]{tuning};
\draw (LO_short) to (2,0.8) node[mslstub, mstlinelen=1.2, rotate=90]{$\lambda_{LO}/4$};

\draw (LO_short) to (2.5,-0.5) node[mslstub, mstlinelen=0.8, rotate=-45]{tuning};
\draw (LO_short) to (2,-0.8) node[mslstub, mstlinelen=1.2, rotate=-90]{$\lambda_{LO}/4$};

\draw (LO_short) to[short] (2.3,0)
      to[mstline, mstlinelen=0.8] (3.5,0)
      to (3.8,0) node[circ](if){};
\draw (if) to[short] (4.1,0) to[mstline, mstlinelen=0.8] (5.5,0) node[msport]{RF};

\draw (if) to[short,o-] ++(0,-1)
to[bandstop,>] ++(0,-2)
to[short] ++(0.3,0) node[msport]{IF};

\draw[rounded corners=0.4cm, thick, orange] (5.45,2) -- (5.45,-1) -- (6.5,-1) -- (6.5,2); 
\node at (6,+1) {RF guide};

\end{circuitikz}

%% file: Chapter/results.tex
\section{Results}
\begin{figure}[H]
    \centering
    \input{figs/Fig5_Trec_600}
    \caption{Receiver noise temperature. Measured and simulated receiver noise temperature (DSB) of the 600-GHz mixer versus pump frequency ($2 \times f_{LO}$). The Y-factor measurement was performed at room temperature, with a relative humidity 32\%, and with a local oscillator power of approximately 2~mW \cite{Jacob2015}. Data are corrected for atmospheric transmission loss of 16~cm.}
    \label{fig:noise600}
\end{figure}
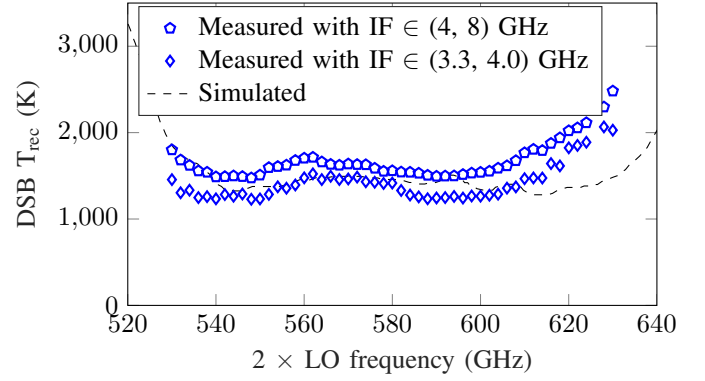

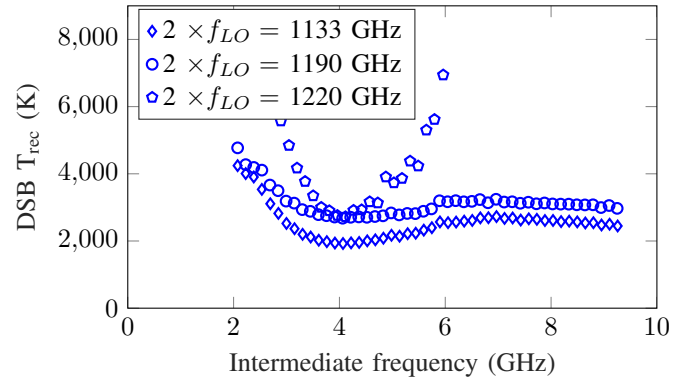
\begin{figure}[H]
    \centering
    \input{figs/Fig6_Trec_1200}
    \caption{Receiver noise temperature. Measured double-sideband (DSB) receiver noise temperature of the 1200-GHz mixer versus intermediate frequency (IF). The Y-factor measurement was carried out at room temperature, with a relative humidity 17\% and a local oscillator power of about 2~mW. Data are corrected for atmospheric transmission loss over an optical path length of 1.1~m \cite{Kochanov2016}.}
    \label{fig:noise1200}
\end{figure}

The measured double-sideband (DSB) receiver noise temperatures for the 600-GHz and 1200-GHz mixers are shown in Figs. \ref{fig:noise600} and \ref{fig:noise1200}, respectively. The receiver noise measurements were corrected for atmospheric loss using the \mbox{HITRAN} molecular absorption database, applying the correction with the lowest sideband attenuation \cite{Gordon2017, Kochanov2016}. Because of the high optical-path loss, the 1200-GHz mixer was characterized within three atmospheric windows as a function of IF frequency, corresponding to $2 \times f_{LO} = $ 1133 1190, and 1220 GHz; see Fig. \ref{fig:noise1200}. As summarized in Table \ref{table_SoA}, the measured receiver noise temperatures are comparable to the best reported room-temperature subharmonic Schottky receivers in the corresponding frequency ranges. The broadband performance of the proposed topology is evident from the results in Fig. \ref{fig:noise600}. Moreover, the measured frequency dependence agrees well with the simulated response, particularly over the central part of the operating band. However, a drawback of this topology is the relatively long transmission line on the RF side, which causes additional loss. 

\begin{table*}[ht]
\renewcommand{\arraystretch}{1.3}
\caption{Comparison of subharmonic Schottky-barrier diode receivers at room temperature.}
\label{table_SoA}
\begin{tabularx}{\textwidth}{@{}c c C c@{}}
\toprule
\makecell{Frequency \\ (GHz)} & \makecell{Minimum DSB receiver noise temperature \\ (K)} & \makecell{Technology} & \makecell{Reference} \\
\midrule
1100-1260 & 2\,800 & JPL, monolithic integration, antiparallel GaAs SBD pair & \cite{Schlecht2014} \\ 
1080-1280 &  3\,000 & LERMA, monolithic integration, antiparallel GaAs SBD pair & \cite{Treuttel2023} \\ 
1130-1220 &  2\,000 & Chalmers, monolithic integration, antiparallel GaAs SBD pair & This work \\ 
874 &  3000 & Chalmers, monolithic integration, antiparallel GaAs SBD pair & \cite{Hammar2018} \\
664 &  1\,800 & RPG, discrete, antiparallel GaAs SBD pair & \cite{Thomas2023} \\
557 &  1300 & Chalmers, monolithic integration, antiparallel GaAs SBD pair & \cite{HuanZhao2012} \\
520-620 &  1130 & LERMA, monolithic integration, antiparallel GaAs SBD pair & \cite{Treuttel2016} \\
520-590 &  3\,000  & JPL, monolithic integration, biasable, shunt antiparallel GaAs SBD pair & \cite{Schlecht2007} \\
520-630 &  1230 & Chalmers, monolithic integration, antiparallel GaAs SBD pair & This work \\

\bottomrule
\end{tabularx}
\end{table*}

The mixer topology shown was also used in the airborne ice cloud imager (ISMAR) operating at 874~GHz \cite{Hammar2018} and in polarimetric receivers at 325~GHz, 424~GHz, and 650~GHz \cite{Auriacombe2026}. The 600-GHz mixer has been further characterized \cite{Jacob2019}, qualified, and delivered by Omnisys Instruments as a flight module for the SWI instrument on board ESA's Juice mission, which is expected to reach the Jupiter system in 2031. LERMA-LPN provided the 1200-GHz receiver flight module for Juice/SWI \cite{Treuttel2023}. In 2024, the overall performance of the submillimeter-wave instrument was tested during a close flyby of the Moon and Earth, both serving as calibration and science targets, demonstrating sufficient stability and data quality for its future endeavors \cite{Hartogh2026, Cavallie2026, Jarchow2026}.

\section{Conclusions}
A compact broadband subharmonic mixer topology has been demonstrated at 600 GHz and 1200 GHz. By placing the RF/LO impedance-matching and virtual-ground structures in close proximity to the antiparallel Schottky diode pair, the electrical length of the critical matching circuitry is reduced. The 600-GHz mixer demonstrates approximately 20\% RF bandwidth with a minimum DSB receiver noise temperature of 1230 K at 590 GHz, while the 1200-GHz implementation achieves a minimum DSB receiver noise temperature of approximately 2000 K around 1130 GHz. These results confirm both broadband operation and the scalability of the proposed circuit topology.

A key feature of the approach is that the critical matching elements are defined lithographically as part of the integrated diode circuit, reducing their dependence on waveguide and assembly tolerances. The main limitation is the relatively long planar RF transmission line, which introduces additional loss and consequently degrades the receiver noise temperature.

Overall, the demonstrated performance at two widely separated frequency bands shows that the proposed topology provides a practical approach to broadband, room-temperature submillimeter-wave receivers and is particularly attractive for spaceborne and remote-sensing instrumentation.

%% file: figs/Fig5_Trec_600.tex
%
%

\begin{tikzpicture}

\begin{axis}[%
width=7cm,
height=4cm,
at={(0.982in,0.627in)},
scale only axis,
xmode=normal,
xmin=520,
xmax=640,
xminorticks=true,
xmajorticks=true,
xlabel style={font=\color{white!15!black}},
xlabel={2 $\times$ LO frequency (GHz)},
ymode=normal,
ymin=0,
ymax=3500,
yminorticks=true,
ylabel style={font=\color{white!15!black}},
ylabel={DSB T$_{\text{rec}}$ (K)},
axis background/.style={fill=white},
legend style={legend cell align=left, align=left, draw=white!15!black, at={(0.02,0.99)},anchor=north west},
]

\addplot [color=blue, only marks, mark=pentagon, mark options={thick}, mark size = 2pt]
  table[row sep=crcr]{%
530.016	1800.932642\\
531.984	1682.866787\\
534	1621.110046\\
536.016	1556.529854\\
537.984	1541.214158\\
540	1487.748134\\
542.016	1490.563327\\
543.984	1498.83735\\
546	1491.360148\\
548.016	1475.399126\\
549.984	1508.178245\\
552	1595.215194\\
554.016	1607.581604\\
555.984	1623.561995\\
558	1677.712333\\
560.016	1704.172863\\
561.984	1714.202278\\
564	1660.679503\\
566.016	1632.474545\\
567.984	1622.603506\\
570	1636.801927\\
572.016	1630.4754\\
573.984	1628.134812\\
576	1587.45498\\
578.016	1552.134308\\
579.984	1558.489555\\
582	1543.494456\\
584.016	1541.125339\\
585.984	1529.788609\\
588	1506.118576\\
590.016	1494.552036\\
591.984	1497.249936\\
594	1499.011797\\
596.016	1515.056843\\
597.984	1531.585659\\
600	1540.36702\\
602.016	1553.560272\\
603.984	1588.298313\\
606	1616.076855\\
608.016	1675.392739\\
609.984	1766.667847\\
612	1806.661398\\
614.016	1792.564705\\
615.984	1875.419657\\
618	1939.631421\\
620.016	2021.646285\\
621.984	2054.895923\\
624	2114.154873\\
627.984	2296.487801\\
630	2481.207446\\
};
\addlegendentry{Measured with IF $\in$ (4, 8) GHz}

\addplot [color=blue, only marks, mark=diamond, mark options={thick}, mark size = 2pt]
  table[row sep=crcr]{%
530.016		1456.053992\\
531.984		1303.737201\\
534		1333.52868\\
536.016		1247.96235\\
537.984		1258.912525\\
540		1233.340949\\
542.016		1282.970978\\
543.984		1263.703997\\
546		1289.448214\\
548.016		1228.614673\\
549.984		1232.292319\\
552		1284.986971\\
554.016		1372.702104\\
555.984		1356.41353\\
558		1393.877234\\
560.016		1476.37978\\
561.984		1522.711456\\
564		1456.429938\\
566.016		1494.270026\\
567.984		1457.409637\\
570		1463.454881\\
572.016		1485.292876\\
573.984		1430.981815\\
576		1425.469317\\
578.016		1411.326621\\
579.984		1415.919679\\
582		1330.41127\\
584.016		1277.53021\\
585.984		1255.892873\\
588		1234.469311\\
590.016		1243.313255\\
591.984		1248.200538\\
594		1261.275752\\
596.016		1242.47557\\
597.984		1266.477369\\
600		1264.371562\\
602.016		1275.217138\\
603.984		1287.559475\\
606		1357.33053\\
608.016		1371.640248\\
609.984		1466.827193\\
612		1475.242271\\
614.016		1474.369911\\
615.984		1641.695249\\
618		1614.609849\\
620.016		1823.812028\\
621.984		1851.771149\\
624		1888.708203\\
627.984		2066.003218\\
630		2028.123925\\
};
\addlegendentry{Measured with IF $\in$ (3.3, 4.0) GHz}

\addplot [color=black, dashed]
  table[row sep=crcr]{%
520.0560204	3258.258981\\
520.6062208	3189.276156\\
521.1564211	3103.068403\\
521.7066215	3034.085577\\
522.1192718	2965.12353\\
522.531922	2913.386411\\
523.0821224	2844.403586\\
523.4947727	2775.441539\\
523.7698729	2723.725198\\
524.1825231	2654.763151\\
524.4576233	2585.821882\\
525.0078237	2499.614128\\
525.558024	2396.181446\\
525.9706743	2327.219399\\
526.3833246	2258.257352\\
526.933525	2189.274527\\
527.4837253	2103.066773\\
528.0339257	2051.308876\\
528.5841261	1982.326051\\
529.1343264	1930.568154\\
529.6845268	1878.810257\\
530.3722772	1827.031583\\
531.0600277	1792.477836\\
531.6102281	1757.944868\\
532.2979785	1740.61605\\
532.985729	1706.062303\\
533.8110295	1688.712707\\
534.49878	1671.383889\\
535.1865304	1654.055071\\
535.8742809	1636.726253\\
536.5620313	1602.172506\\
537.2497818	1584.843688\\
537.7999822	1550.31072\\
538.6252827	1515.736195\\
539.3130332	1463.95752\\
539.8632335	1429.424552\\
540.550984	1394.870805\\
541.2387344	1377.541987\\
541.9264849	1342.988241\\
542.6142354	1342.884351\\
543.3019858	1342.780461\\
543.9897363	1325.451643\\
544.8150368	1325.326975\\
545.6403374	1325.202307\\
546.4656379	1342.302568\\
547.4284885	1359.382051\\
548.2537891	1376.482311\\
549.2166397	1376.336866\\
550.1794904	1376.19142\\
551.0047909	1376.066752\\
551.8300915	1375.942084\\
552.655392	1375.817417\\
553.6182426	1375.671971\\
554.5810933	1375.526525\\
555.5439439	1392.606008\\
556.9194448	1409.623156\\
558.4324958	1443.844456\\
559.3953465	1460.923938\\
560.7708474	1443.49123\\
562.6965487	1477.650196\\
563.3842991	1477.546306\\
565.0349002	1477.29697\\
566.8230514	1494.251785\\
568.7487527	1493.960894\\
569.4365031	1493.857004\\
570.9495541	1493.628446\\
572.8752554	1493.337555\\
574.8009567	1475.821735\\
576.726658	1475.530844\\
577.5519585	1458.181248\\
578.9274594	1457.973468\\
580.7156106	1457.703355\\
581.8160113	1457.537131\\
582.778862	1440.166757\\
583.8792627	1422.775605\\
584.9796634	1422.609381\\
586.2176142	1405.197451\\
586.2176142	1405.197451\\
587.318015	1405.031227\\
588.6935159	1422.048376\\
589.7939166	1421.882153\\
590.7567673	1438.961635\\
591.4445177	1473.307602\\
592.1322682	1490.428641\\
593.2326689	1507.487345\\
593.9204194	1507.383455\\
594.88327	1490.013081\\
595.5710204	1472.684263\\
596.396321	1420.88481\\
597.0840714	1420.780921\\
597.6342718	1386.247952\\
598.3220223	1386.144062\\
599.1473228	1351.569538\\
600.6603738	1334.116052\\
601.4856744	1333.991384\\
602.7236252	1333.804383\\
603.4113756	1368.15035\\
604.6493265	1385.188276\\
605.474627	1402.288537\\
606.8501279	1402.080757\\
607.5378784	1401.976867\\
608.9133793	1384.54416\\
609.7386798	1332.744707\\
610.2888802	1315.436667\\
611.2517308	1298.066293\\
612.2145815	1280.695918\\
613.3149822	1280.529695\\
614.4153829	1280.363471\\
615.2406835	1280.238803\\
616.3410842	1297.297508\\
617.4414849	1331.581141\\
618.5418857	1348.639846\\
619.6422864	1365.69855\\
620.7426871	1365.532327\\
621.7055378	1365.386881\\
622.6683884	1365.241435\\
623.493689	1382.341696\\
624.0438893	1382.258584\\
625.00674	1382.113138\\
625.5569403	1399.254955\\
626.519791	1399.109509\\
627.4826416	1433.41392\\
628.3079421	1450.514181\\
629.2707928	1467.593663\\
630.0960933	1484.693924\\
631.058944	1501.773406\\
631.7466944	1536.119373\\
632.8470952	1570.403006\\
633.6723957	1621.953123\\
634.9103465	1673.440907\\
636.0107472	1724.949468\\
636.9735979	1776.478807\\
637.7988984	1828.028925\\
638.3490988	1879.620598\\
639.3119494	1965.599794\\
640.13725	2034.374839\\
};
\addlegendentry{Simulated}

\end{axis}
\end{tikzpicture}

%% file: figs/Fig6_Trec_1200.tex
%
%

\begin{tikzpicture}

\begin{axis}[%
width=7cm,
height=4cm,
at={(0.982in,0.627in)},
scale only axis,
xmode=normal,
xmin=0,
xmax=10,
xminorticks=true,
xmajorticks=true,
xlabel style={font=\color{white!15!black}},
xlabel={Intermediate frequency (GHz)},
ymode=normal,
ymin=0,
ymax=9000,
yminorticks=true,
ylabel style={font=\color{white!15!black}},
ylabel={DSB T$_{\text{rec}}$ (K)},
axis background/.style={fill=white},
legend style={legend cell align=left, align=left, draw=white!15!black, at={(0.02,0.99)},anchor=north west},
]

\addplot [color=blue, only marks, mark=diamond, mark options={thick}, mark size = 2pt]
  table[row sep=crcr]{%
2.07617	4242.572 \\
2.22876	4006.86 \\
2.38135	3904.45 \\
2.53394	3535.732 \\
2.69385	3108.355 \\
2.84644	2821.593 \\
2.99902	2522.423\\
3.15161	2365.871\\
3.3042	2198.122\\
3.45679	2117.555\\
3.60968	2023.159\\
3.76227	1980.334\\
3.91486	1938.827\\
4.06775	1924.974\\
4.22034	1943.954\\
4.37292	1956.463\\
4.52612	2004.562\\
4.67871	2034.541\\
4.8313	2082.75\\
4.98389	2170.288\\
5.13647	2142.317\\
5.28937	2217.336\\
5.44196	2231.365\\
5.59454	2322.729\\
5.74713	2394.933\\
5.89972	2567.259\\
6.05267	2545.303\\
6.20526	2563.411\\
6.35785	2590.961\\
6.51044	2611.877\\
6.66302	2688.272\\
6.81561	2692.688\\
6.9682	2726.262\\
7.12079	2671.889\\
7.27368	2684.498\\
7.42627	2622.805\\
7.57886	2659.834\\
7.73145	2644.786\\
7.88403	2609.366\\
8.03693	2610.872\\
8.18951	2577.223\\
8.3421	2591.091\\
8.49469	2568.093\\
8.64728	2536.39\\
8.79987	2543.7\\
8.95306	2476.363\\
9.10565	2498.447\\
9.25824	2447.077\\
};
\addlegendentry{2 $\times f_{LO}=$ 1133 GHz}

\addplot [color=blue, only marks, mark=o, mark options={thick}, mark size = 2pt]
  table[row sep=crcr]{%
2.07617	4772.515\\
2.22876	4273.429\\
2.38135	4189.665\\
2.53394	4109.529\\
2.68683	3665.137\\
2.83972	3497.654\\
2.99231	3187.004\\
3.1449	3127.639\\
3.29749	2933.887\\
3.45007	2883.253\\
3.60266	2784.315\\
3.75525	2753.042\\
3.90784	2712.999\\
4.06073	2678.363\\
4.21332	2706.116\\
4.36591	2716.646\\
4.51849	2705.781\\
4.67108	2728.161\\
4.82367	2749.121\\
4.97626	2835.101\\
5.12885	2776.348\\
5.28143	2820.096\\
5.43402	2814.957\\
5.58661	2885.026\\
5.7392	2955.212\\
5.89178	3197.234\\
6.04443	3168.214\\
6.19702	3200.38\\
6.34961	3166.305\\
6.5022	3179.81\\
6.65479	3232.724\\
6.80737	3138.466\\
6.95996	3241.744\\
7.11255	3159.516\\
7.26514	3167.465\\
7.41772	3123.333\\
7.57031	3153.934\\
7.7229	3107.557\\
7.87549	3131.832\\
8.02838	3102.975\\
8.18097	3096.975\\
8.33356	3096.242\\
8.48615	3077.772\\
8.63873	3073.808\\
8.79132	3077.502\\
8.94391	2995.436\\
9.0965	3055.824\\
9.24908	2971.484\\
};
\addlegendentry{2 $\times f_{LO}=$ 1190 GHz}

\addplot [color=blue, only marks, mark=pentagon, mark options={thick}, mark size = 2pt]
  table[row sep=crcr]{%
2.08319	13150.42\\
2.26569	11539.75\\
2.42651	9156.661\\
2.58398	7822.507\\
2.74023	7139.47\\
2.89313	5573.246\\
3.04602	4844.324\\
3.19861	4170.31\\
3.3512	3771.812\\
3.50378	3345\\
3.65637	3000.883\\
3.80896	2900.8\\
3.96155	2759.734\\
4.11444	2720.72\\
4.26703	2912.459\\
4.41962	2927.96\\
4.5722	3166.259\\
4.72479	3132.525\\
4.87738	3905.465\\
5.02997	3731.38\\
5.18256	3862.857\\
5.33545	4377.665\\
5.48834	4232.333\\
5.64246	5300.874\\
5.79626	5617.835\\
5.96106	6941.734\\
};
\addlegendentry{2 $\times f_{LO}=$ 1220 GHz}

\end{axis}
\end{tikzpicture}